\documentclass[amsmath,amsfonts,amssymb,twocolumn,nofootinbib]{revtex4-1}
\usepackage{graphicx}
\usepackage{verbatim}
\usepackage{hyperref}

\begin{document}

\title{Cruising The Simplex: Hamiltonian Monte Carlo and the Dirichlet Distribution}
\author{Michael Betancourt}
\affiliation{Massachusetts Institute of Technology, Cambridge, MA 02139}
\email{betanalpha@gmail.com}

\begin{abstract}
Due to its constrained support, the Dirichlet distribution is uniquely suited to many applications.  The constraints that make it powerful, however, can also hinder practical implementations, particularly those utilizing Markov Chain Monte Carlo (MCMC) techniques such as Hamiltonian Monte Carlo.  I introduce a series of transformations that reshape the canonical Dirichlet distribution into a form more amenable to efficient MCMC algorithms and demonstrate its utility with a few examples.
\end{abstract}


\maketitle

\section*{The Dirichlet Distribution}

Given $m$ random variables, $\mathbf{x}$, with the constrained support
\begin{equation*}
\begin{array}{c}
0 \leq x_{i} \leq 1 \\
\displaystyle\sum\limits_{i = 1}^{m} x_{i}= 1 \\
\end{array},
\end{equation*}
the Dirichlet distribution \cite{MacKay2003, Bishop2007} is defined by the parameterized probability density
\begin{equation*}
\mathrm{Dir} \left( \mathbf{x} | \mbox{\boldmath{$\alpha$}} \right) = \frac{\Gamma \left( \sum_{i = 1}^{m} \alpha_{i} \right) }{\prod_{i = 1}^{m} \Gamma \left( \alpha_{i} \right) } \cdot \prod_{i = 1}^{m} x_{i}^{\alpha_{i} - 1}, 
\end{equation*}
where $\alpha_{i} \in \mathbb{R}^{+}$  and $\Gamma$ is the usual gamma function.
Because of its distinctive support, the Dirichlet distribution is particularly well-suited for modeling the allocation of conserved quantities such as probability.  The distribution becomes invaluable when studying categorical problems: inference with histograms a pervasive example.

\subsection*{Generating Dirichlet Samples}

Sampling directly from the Dirichlet distribution is made feasible due to a convenient property of the Gamma distribution \cite{Cheng1998}.  An ensemble of Gamma variates,
\begin{equation*}
u_{i} \sim \mathrm{Ga} \left( u_{i} |  \alpha_{i}, \beta_{i} \right),
\end{equation*}
follows a Dirichlet distribution upon normalization,
\begin{equation*}
\left\{ \frac{ u_{i} }{ \sum_{j} u_{j} } \right\} = \left\{ x_{i} \right\} \sim \mathrm{Dir} \left( \mathbf{x} | \mbox{\boldmath{$\alpha$}} \right).
\end{equation*}
Given the efficiency of modern Gamma generators \cite{Marsaglia2000}, the generation of independent Dirichlet variates is not particularly demanding.

The same property admits a parallel Markov Chain Monte Carlo approach.  Here separate chains generate the independent Gamma variates, which are then normalized to produce the desired Dirichlet sample.  Because the chains interact only in the normalization, however, their evolutions are uncorrelated and consequently useless to algorithms that rely on the local correlations of the distribution such as nested sampling \cite{Skilling2004}.  Moreover, the need to interrupt the evolution to normalize prohibits the generalization to chains sampling from a posterior distributions predicated on a Dirichlet prior.

Creating a Markov chain that samples directly from the Dirichlet distribution is complicated by the constraints, as the Dirichlet variates manifest not in $m$ dimensions but rather in a $m - 1$ dimensional submanifold known as a simplex.  Unless the Markov transitions accommodate the constraints directly, proposed samples will fall outside of the simplex (with probability 1) and the chain will be unable to progress beyond an initial seed.

By parameterizing the simplex directly, the constraints are implicitly taken into account and the chain will have no problem exploring the full support of the Dirichlet distribution.  Constructing a systematic map between the original $m$ dimensional manifold and the simplex is straightforward, if a bit ungainly, but real difficulties begin to arise when considering the boundary of the simplex.  

Because the simplex is the inclusive volume of a $m - 1$ dimensional polytope, the support is bounded by a surface of piecewise faces, the number of which grows exponentially with the dimension of the original distribution.  Modeling this complex boundary in terms of simplex coordinates is awkward at best, especially when appealing to more sophisticated algorithms like constrained Hamiltonian Monte Carlo \cite{Neal2011, Betancourt2011} that require a careful description of the boundary geometry.

\section*{Smoothing Out The Simplex}

A simple change of variables dramatically simplifies the structure of the simplex.  Taking
\begin{equation*}
y_{i} = \sqrt{x_{i}},
\end{equation*}
the original Dirichlet distribution becomes
\begin{equation*}
\mathrm{Dir} \left( \mathbf{y} | \mbox{\boldmath{$\alpha$}} \right) = 2^{m} \frac{\Gamma \left( \sum_{i = 1}^{m} \alpha_{i} \right) }{\prod_{i = 1}^{m} \Gamma \left( \alpha_{i} \right) } \cdot \prod_{i = 1}^{m} y_{i}^{2 \alpha_{i} - 1}
\end{equation*}
with the support
\begin{equation*}
\begin{array}{c}
0 \leq y_{i} \leq 1 \\
\displaystyle\sum\limits_{i = 1}^{m} y_{i}^{2} = 1 \\
\end{array}.
\end{equation*}
The quadratic constraint defines a substantially simpler submanifold: the surface of an $m$ dimensional hypersphere within the positive orthant.

The surface of the hypersphere can be parameterized directly by transforming to hyperspherical coordinates \cite{Hassani2002},
\begin{equation*}
y_{i} = r \left( \prod_{k = 1}^{i - 1} \sin \theta_{k} \right) \cdot \left\{ \begin{array}{rc} \cos \theta_{i} ,& i < m \\ 1 ,& i = m \end{array} \right. .
\end{equation*}
In these coordinates the constrained support becomes
\begin{equation*}
\begin{array}{c}
0 \leq \theta_{i} \leq \dfrac{\pi}{2} \\
r^{2} = 1 \\
\end{array}
\end{equation*}
with the distribution
\begin{align*}
\mathrm{Dir} \left( r, \mbox{\boldmath{$\theta$}} | \mbox{\boldmath{$\alpha$}} \right) =& \, 2^{m} \frac{\Gamma \left( \sum_{i = 1}^{m} \alpha_{i} \right) }{\prod_{i = 1}^{m} \Gamma \left( \alpha_{i} \right) } \cdot r^{2 \widetilde{\alpha}_{0} - 1} \\
& \times \prod_{i = 1}^{m - 1} \left( \cos \theta_{i} \right)^{2 \alpha_{i} - 1} \left( \sin \theta_{i} \right)^{2 \widetilde{\alpha}_{i} - 1},
\end{align*}
where
\begin{equation*}
\widetilde{\alpha}_{i} = \sum_{k = i + 1}^{m} \alpha_{k}.
\end{equation*}

Marginalization over the radial coordinate is immediate due to the constraint, giving a distribution for the hyperspherical angles alone
\begin{align*}
\mathrm{Dir} \left( \mbox{\boldmath{$\theta$}} | \mbox{\boldmath{$\alpha$}} \right) =& \, 2^{m - 1} \frac{\Gamma \left( \sum_{i = 1}^{m} \alpha_{i} \right) }{\prod_{i = 1}^{m} \Gamma \left( \alpha_{i} \right) } \\
& \prod_{i = 1}^{m - 1} \left( \cos \theta_{i} \right)^{2 \alpha_{i} - 1} \left( \sin \theta_{i} \right)^{2 \widetilde{\alpha}_{i} - 1}, \\
& 0 \leq \theta_{i} \leq \frac{\pi}{2}.
\end{align*}

The abundance of trigonometric functions is a bit awkward, but substituting 
\begin{equation*}
z_{i} = \sin^{2} \theta_{i}
\end{equation*}
gives 
\begin{equation*}
\mathrm{Dir} \left( \mathbf{z} | \mbox{\boldmath{$\alpha$}} \right) = \prod_{i = 1}^{m - 1} \mathrm{Be} \left( \widetilde{\alpha}_{i}, \alpha_{i} \right),
\end{equation*}
where 
\begin{equation*}
0 \leq z_{i} \leq 1.
\end{equation*}

The combined mapping $\mathbf{x} \rightarrow \mathbf{y} \rightarrow \mbox{\boldmath{$\theta$}} \rightarrow \mathbf{z}$ reduces the original Dirichlet distribution to a simple product of independent Beta distributions and, despite the complexity of the coordinate transformations, the inverse of the combined transformation back to the original manifold is simply 
\begin{equation*}
x_{i} = \left( \prod_{k = 1}^{i - 1} z_{k} \right)  \cdot \left\{ \begin{array}{rc} 1 - z_{i}, & i < m \\ 1 ,& i = m \end{array} \right. .
\end{equation*}
All of the computationally expensive function evaluations have cancelled, requiring only simple operations to transform between the two spaces.

By warping the original manifold, the complicated correlations of the Dirichlet distribution have simplified dramatically.  The simplex is gone, replaced by a product of univariate distributions particularly accommodating to constrained Hamiltonian Monte Carlo.

\subsection*{Differentiating in the Transformed Space}

The ability to run a Markov chain that samples directly from the Dirichlet distribution enbales algorithms such as nested sampling, or extending the chain to sample from a posterior based on a Dirichlet prior with Hamiltonian Monte Carlo.  These applications, however, require gradients with respect to the coordinates of the submanifold $\mathbf{z}$.

Appealing to the chain rule,
\begin{equation*}
\frac{ \partial f \left( \mathbf{x} \right) }{ \partial z_{i} } = \sum_{j= 1}^{m} \frac{ \partial f \left( \mathbf{x} \right) }{ \partial x_{j} } \frac{ \partial x_{j} }{ \partial z_{i} } .
\end{equation*}
Substituting the coordinate derivatives,
\begin{equation*}
\frac{\partial x_{j} }{ \partial z_{i} } = \left\{ \begin{array}{rc} 0 ,& i > j \\ x_{j} / \left( z_{j} - 1 \right),& i = j \\ x_{j} / z_{i},& i < j \\ \end{array} \right. ,
\end{equation*}
gives 
\begin{equation*}
\frac{ \partial f \left( \mathbf{x} \right) }{ \partial z_{i} } =  \left( \frac{x_{i}}{z_{i} - 1} \right) \frac{ \partial f \left( \mathbf{x} \right) }{ \partial x_{i} } + \sum_{j = i + 1}^{m} \frac{x_{j}}{z_{i}} \frac{ \partial f \left( \mathbf{x} \right) }{ \partial x_{j} } .
\end{equation*}

Once the original gradient, $\partial f / \partial x_{i}$, has been computed the new gradient can be found with only a few additional multiplications and divisions.

\section*{Examples}

Given the ubiquity of the Dirichlet distribution in modern inference problems there is no shortage of interesting examples, but here I mention only two: fitting histograms and deconvolution.  Both take advantage of the above transformation to sample from the posteriors of many variables with Hamiltonian Monte Carlo, ultimately marginalizing out nuisance parameters not desired in the final inference.

\subsection*{Fitting Histograms}

Histograms are popular in many fields due to their ease of use and effectiveness as non-parametric models.  When histograms are combined into larger analyses, however, their uncertainties can be difficult to incorporate.  

Consider, for example, fitting a measured histogram to the sum of two spectra (such as signal and background) represented with normalized histograms.  If the latter are known exactly, then the likelihood of the measured histogram is simply the product of independent Poisson distributions, with source strength given by the weighted sum of the two spectra, and any sort of fit is straightforward.  In practice, however, those spectra often come from computationally expensive simulations or time consuming calibrations and the uncertainties are not only nontrivial but can dominate the sampling uncertainty of the measured histogram.

Modeling the normalized histograms with Dirichlet priors admits a posterior over both the spectra and their weights, with the bin contents then marginalized out to give a posterior for the weights alone that fully incorporates all sources of uncertainty in the problem (Fig \ref{fig:shapes} and \ref{fig:postShape}).
 
\subsection*{Deconvolution}

Common to many analyses, deconvolution is a notoriously difficult problem.  One obstacle is that many algorithms fail to constrain the number of events, which must be preserved in a linear convolution, or take into account any uncertainties in the convolution function.  When the data are binned, both the deconvolved histogram and the columns of the convolution matrix are modeled with Dirichlet priors and the entire system can be handled at once, sampling for the full posterior over all uncertain values and marginalizing over the entries of the convolution matrix to give the final posterior.

Extensions to this general strategy allow continuity constraints and even non-square convolution matrices, furnishing a comprehensive algorithm that can handle a range of difficult problems without active tuning (Fig \ref{fig:decoExp} and \ref{fig:decoSumGauss}).

\section*{Acknowledgements}

I thank Kat Deck, John Rutherford, Leo Stein, and Phil Zukin for helpful discussion and comments.

\bibliography{dirichletHMC}

\pagebreak

\begin{widetext}

\begin{figure}[h]
\centering
\includegraphics[width=5.9in]{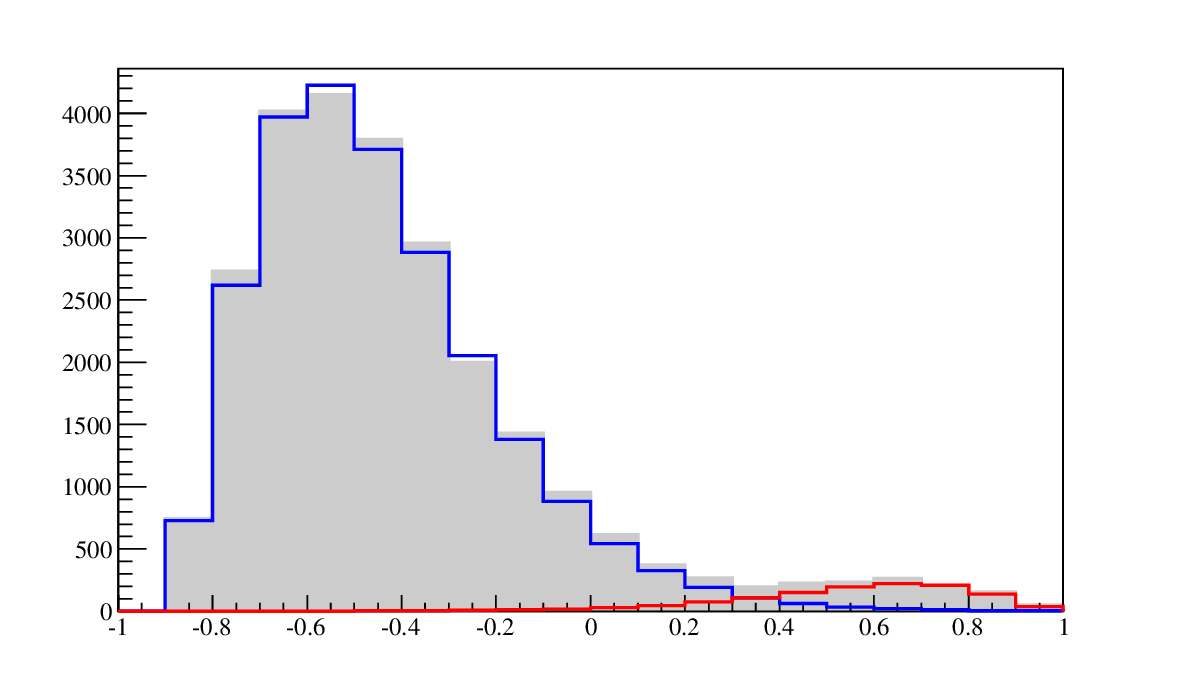}
\caption{By modeling the summed spectra (red and blue) with Dirichlet priors, their uncertain values can be marginalized out of the fit to the data (gray), yielding a posterior for the normalizations alone.
\label{fig:shapes} }
\end{figure}

\begin{figure}[h]
\centering
\includegraphics[width=5.9in]{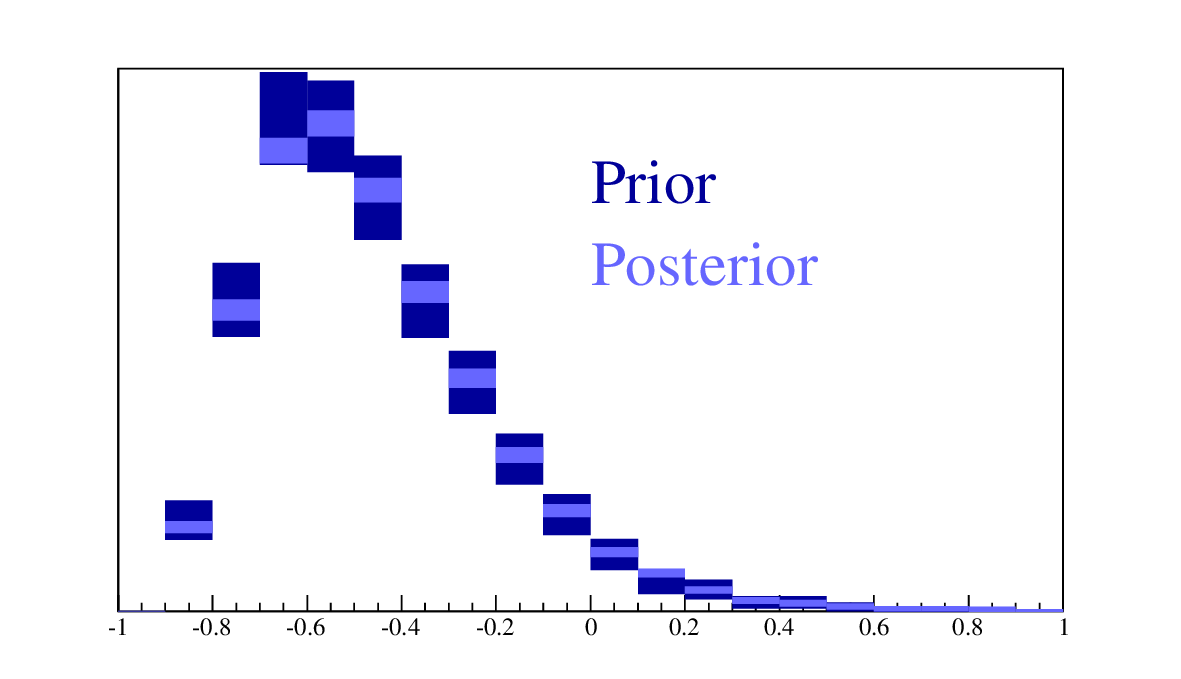}
\caption{When the data is known more precisely than the initial spectra, incorporating the initial spectra into the posterior with Dirichlet priors provides a significant reduction in their uncertainty, as seen in the 95\% posterior intervals for the blue spectrum from Figure \ref{fig:shapes}.
\label{fig:postShape} }
\end{figure}

\pagebreak

\begin{figure}[h]
\centering
\includegraphics[width=7in]{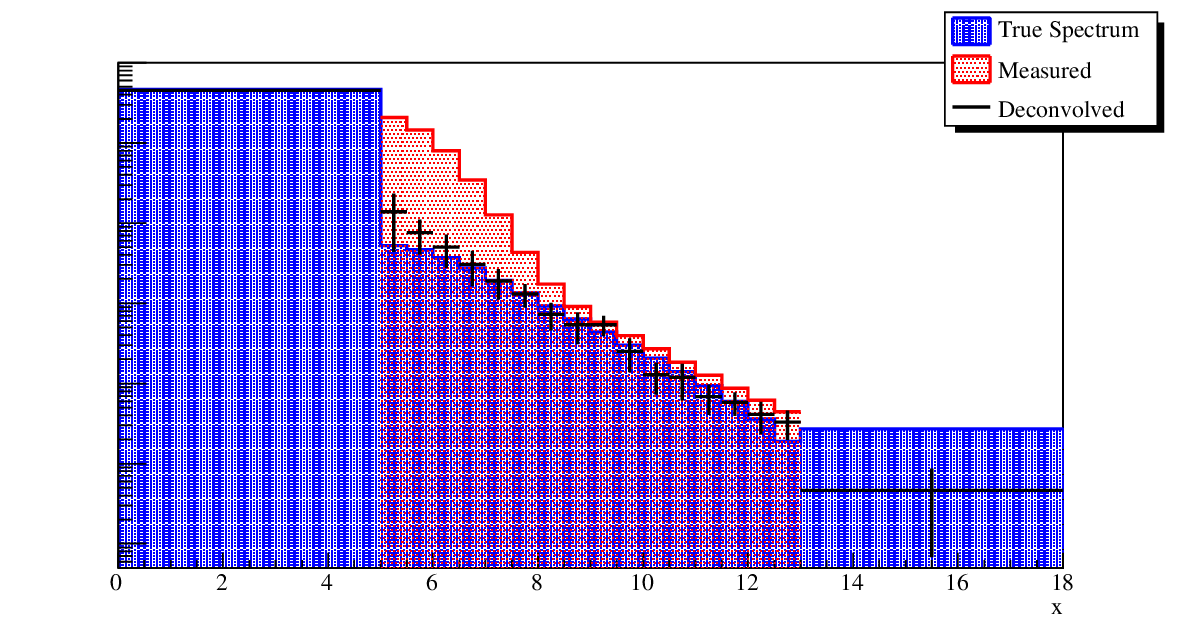}
\caption{Non-square convolution matrices appear when events spill over boundaries in the nominal measurement.  Here deconvolution with extensive use of Dirichlet priors correctly removes the excess events that leaked into the measured spectrum from below the $x = 5$ analysis threshold.
\label{fig:decoExp} }
\end{figure}

\begin{figure}[h]
\centering
\includegraphics[width=7in]{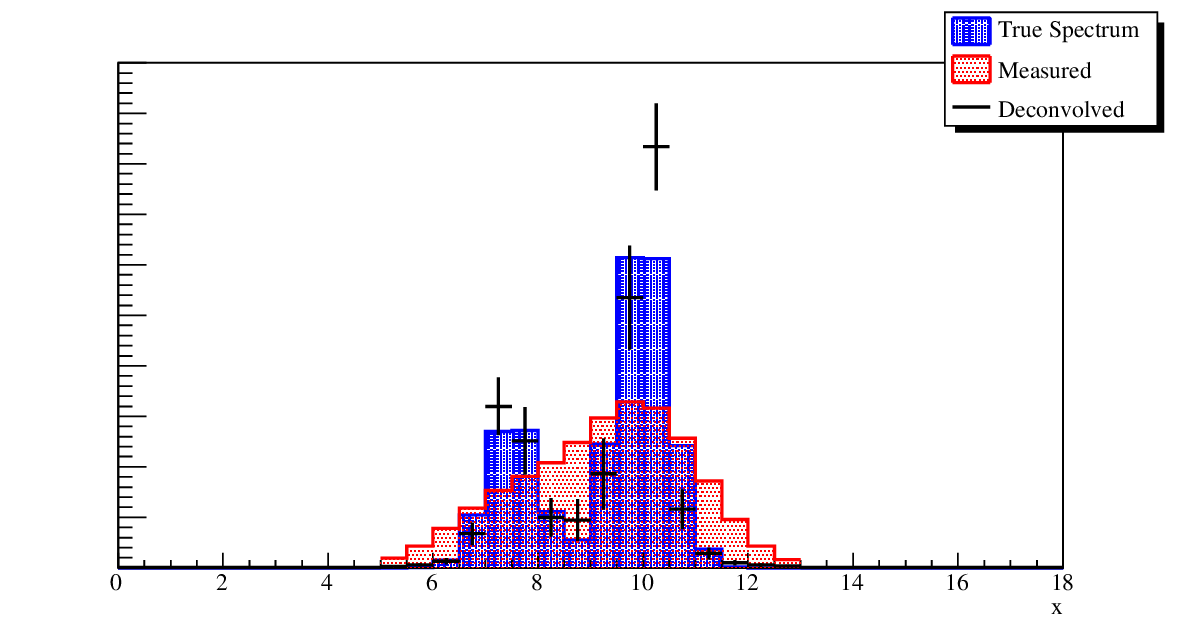}
\caption{With no active tuning, the same algorithm in Figure \ref{fig:decoExp} correctly recovers two peaks despite minimal evidence in the measured spectrum.  Note the dramatic difference in length scales between the two problems.
\label{fig:decoSumGauss} }
\end{figure}

\end{widetext}

\end{document}